\documentclass[11pt]{article}

\usepackage[bookmarks=true]{hyperref}
\hypersetup{breaklinks=true}

\let\fn\footnote
\renewcommand{\footnote}[1]{\linespread{1.1}\fn{#1}\linespread{1.29}}

\usepackage{fancyhdr}
\usepackage[left]{lineno}

\makeatletter\renewcommand{\section}{\@startsection
{section}{1}{\z@}{-3.5ex plus -1ex minus
    -.2ex}{2.3ex plus .2ex}{\bf }}
\makeatletter\renewcommand{\subsection}{\@startsection{subsection}{2}{\z@}{-3.25ex
plus -1ex minus
   -.2ex}{1.5ex plus .2ex}{\it }}
\makeatletter\renewcommand{\subsubsection}{\@startsection{subsubsection}{3}{-2.45ex}{-3.25ex
plus -1ex minus -.2ex}{1.5ex plus .2ex}{\it }}
\renewcommand{\thesection}{\arabic{section}.}
\renewcommand{\thesubsection}{\arabic{section}.\arabic{subsection}.}

\renewcommand{\theequation}{\thesection\arabic{equation}}
\makeatletter \@addtoreset{equation}{section}

\newcommand{\acknowledgements}{\pdfbookmark[1]{Acknowledgements}
{acknowledgements}\section*{Acknowledgements}}

\newcommand{\appendices}{\pdfbookmark[1]{Appendix}{appendices}
\section*{Appendix}\label{appendices}\setcounter{subsection}{0}
\setcounter{equation}{0}
\renewcommand*{\theHequation}{\arabic{equation}}
\renewcommand{\thesubsection}{\Alph{subsection}.}
\renewcommand{\theequation}{\thesubsection\arabic{equation}}
}

\usepackage{epsfig,rotating}
\usepackage{amsmath,amssymb}
\usepackage{amsfonts}
\usepackage{mathrsfs}
\usepackage{bbm}
\usepackage{bm}
\usepackage[vcentermath,enableskew]{youngtab}

\def\slasha#1{\setbox0=\hbox{$#1$}#1\hskip-\wd0\hbox to\wd0{\hss\sl/\/\hss}}

\def\periodb#1{\setbox0=\hbox{$#1$}#1\hskip-\wd0\hbox to\wd0{-}}

\newcommand{\lbr}{(\hspace{-0.1cm}(}
\newcommand{\rbr}{)\hspace{-0.1cm})}
\newcommand{\blbr}{\big(\hspace{-0.1cm}\big(}
\newcommand{\brbr}{\big)\hspace{-0.1cm}\big)}

\newcommand{\unit}{\mathbbm{1}}   			

\newcommand{\CA}{\mathcal{A}}    			

\newcommand{\CCA}{\mathscr{A}}

\newcommand{\CC}{\mathcal{C}}

\newcommand{\CL}{\mathcal{L}}

\newcommand{\CN}{\mathcal{N}}

\newcommand{\CT}{\mathcal{T}}

\newcommand{\CU}{\mathcal{U}}
\newcommand{\CV}{\mathcal{V}}

\newcommand{\CCX}{\mathscr{X}}

\newcommand{\CCY}{\mathscr{Y}}

\newcommand{\CCZ}{\mathscr{Z}}
\newcommand{\CE}{\mathcal{E}}
\newcommand{\frg}{\mathfrak{g}}				

\newcommand{\fru}{\mathfrak{u}}

\newcommand{\FR}{\mathbbm{R}}     			
\newcommand{\FC}{\mathbbm{C}}     			
\newcommand{\NN}{\mathbbm{N}}     			
\newcommand{\CPP}{{\mathbbm{C}P}}    			

\newcommand{\dd}{\mathrm{d}}     			
\newcommand{\dpar}{\partial}     			
\newcommand{\de}{\mathrm{e}}     			
\newcommand{\di}{\mathrm{i}}     			
\newcommand{\eps}{{\varepsilon}}			

\newcommand{\eand}{{~~~~\mbox{and}~~~~}}     		

\newcommand{\tr}{\,\mathrm{tr}\,}     			
\newcommand{\str}{{\,\mathrm{str}\,}}     		

\newcommand{\au}{\mathfrak{u}}
\newcommand{\asu}{\mathfrak{su}}
\newcommand{\aso}{\mathfrak{so}}
\newcommand{\aosp}{\mathfrak{osp}}

\newcommand{\sU}{\mathsf{U}}     			

\newcommand{\sSU}{\mathsf{SU}}

\newcommand{\sSO}{\mathsf{SO}}

\newcommand{\remark}[1]{}     				
\def\tyng(#1){\hbox{\tiny$\yng(#1)$}}			
\def\tyoung(#1){\hbox{\tiny$\young(#1)$}}			

\begin{document}

\begin{titlepage}
\begin{flushright}
 HWM--10--33 \\ EMPG--10--24
\end{flushright}
\vskip 2.0cm
\begin{center}
{\LARGE \bf Quantized Nambu-Poisson Manifolds\\[0.3cm] in a 3-Lie
  Algebra Reduced Model} \vskip 1.5cm
{\Large Joshua DeBellis, Christian S\"amann, Richard J. Szabo}
 \setcounter{footnote}{0}
\renewcommand{\thefootnote}{\arabic{thefootnote}} \vskip 1cm {\em Department of Mathematics\\
Heriot-Watt University\\
Colin Maclaurin Building, Riccarton, Edinburgh EH14 4AS, U.K.\\
and Maxwell Institute for Mathematical Sciences, Edinburgh,
U.K.}\\[5mm] {E-mail: {\ttfamily jd111@hw.ac.uk , C.Saemann@hw.ac.uk , R.J.Szabo@hw.ac.uk}} \vskip 1.1cm
\end{center}
\vskip 1.0cm
\begin{center}
{\bf Abstract}
\end{center}
\begin{quote}
We consider dimensional reduction of the Bagger-Lambert-Gustavsson
theory to a zero-dimensional 3-Lie algebra model and construct various stable
solutions corresponding to quantized Nambu-Poisson manifolds. A
recently proposed Higgs mechanism reduces this model to the IKKT
matrix model. We find that in the strong coupling limit, our solutions
correspond to ordinary noncommutative spaces arising as stable solutions in the IKKT model with D-brane backgrounds. In particular, this
happens for $S^3$, $\FR^3$ and five-dimensional Neveu-Schwarz Hpp-waves. We expand
our model around these backgrounds and find effective noncommutative
field theories with complicated interactions involving
higher-derivative terms. We also describe the relation of our reduced model
to a cubic supermatrix model based on an $\aosp(1|32)$ supersymmetry algebra.
\end{quote}
\end{titlepage}


\section{Introduction}

Dimensional reductions of ten-dimensional maximally supersymmetric Yang-Mills
theory lead to interesting zero-dimensional and one-dimensional matrix
models, called respectively the IKKT~\cite{Ishibashi:1996xs} and
BFSS~\cite{Banks:1996vh} models. The IKKT matrix model is
conjecturally a non-perturbative completion of type~IIB string theory,
while the BFSS matrix quantum mechanics is dual to M-theory in
discrete light-cone quantization on flat space. Their classical
solutions describe brane configurations which have also found interpretations in terms of noncommutative
geometry. For example, expansion of the IKKT matrix model around a
D-brane background preserving part of the supersymmetry yields a supersymmetric gauge theory on Moyal
space~\cite{Aoki:1999vr}, while toroidal compactification of the BFSS
model in a constant $C$-field background gives Yang-Mills theory on a
noncommutative torus~\cite{Connes:1997cr}. The appearance of flat
noncommutative spaces can be understood directly in string theory from
the quantization of open strings ending on D-branes in a constant
$B$-field background~\cite{Seiberg:1999vs}. Studying these equivalences between large $N$ reduced models and noncommutative gauge theories is expected to lead to
new insights as to what extent these matrix models are dual to gauge and
gravitational theories. 

In string theory, fuzzy spheres appear as classical solutions to D0-brane equations of motion in the presence of an external
Ramond-Ramond flux~\cite{Myers:1999ps}. In the IKKT matrix model
description they arise as solutions to the classical equations
of motion if one adds a Chern-Simons term representing the coupling to the external
field~\cite{Iso:2001mg}; expanding the bosonic matrices around the classical solution gives a noncommutative gauge theory on fuzzy spheres.
The corresponding modification of the BFSS model is a massive matrix
model with Chern-Simons term, called the BMN matrix
model~\cite{Berenstein:2002jq}, which conjecturally describes the
discrete light-cone quantization of M-theory on a supersymmetric
pp-wave background and lifts the flat directions of the BFSS model. In this case both fuzzy spheres and fuzzy hyperboloids appear as half-BPS solutions~\cite{Bak:2002rq,Park:2002cba}, and describe static large M2-branes or static large longitudinal M5-branes. 

In this paper we describe an analogous treatment of the Bagger-Lambert-Gustavsson (BLG) theory of multiple M2-branes~\cite{Bagger:2007jr,Gustavsson:2007vu}. We consider a dimensional reduction of this theory to a zero-dimensional 3-Lie algebra model; similar reduced models have also been studied in~\cite{Sato:2009mf,Furuuchi:2009ax,Sato:2010ca,Tomino:2010xw}. One would expect that the noncommutative geometries arising in this context are based on 3-Lie algebras and that they are of the types discussed e.g.\ in~\cite{DeBellis:2010pf}.
We will study the BLG 3-Lie algebra reduced model in detail, and
construct various stable classical solutions corresponding to
noncommutative geometries. The Higgs mechanism proposed recently by
Mukhi and Papageorgakis~\cite{Mukhi:2008ux} connecting the BLG theory
to the maximally supersymmetric Yang-Mills theory in three dimensions
connects here the 3-Lie algebra model to the IKKT matrix model. Using
this mechanism, one can regard the noncommutative geometries
corresponding to stable solutions in the matrix model as strong coupling limits of noncommutative geometries arising in our 3-Lie algebra model. In particular, we find that the fuzzy two-spheres, noncommutative $\FR^2$ and four-dimensional noncommutative Neveu-Schwarz Hpp-waves arise in a strong coupling limit from the fuzzy three-sphere, fuzzy $\FR^3_\lambda$ and five-dimensional noncommutative Hpp-waves, respectively.

We also examine the effective noncommutative gauge theory arising from expanding the action of the 3-Lie algebra model around a stable classical solution corresponding to a noncommutative space. Here we are again confronted with the problem already observed in~\cite{DeBellis:2010pf}: The 3-Lie algebra structure appears only at linear level in the noncommutative geometries. We therefore have to look at possible matrix algebra representations of 3-Lie algebras, which turn the BLG 3-Lie algebra model into a conventional matrix model. The resulting theories are complicated, and do not allow for a straightforward interpretation; it would be interesting to understand their relation to the supersymmetric Yang-Mills matrix quantum mechanics of the BFSS matrix theory which describes M2-branes in light-cone gauge. The one-loop effective action obtained by integrating out linear fluctuations about noncommutative backgrounds in a similar reduced model is considered in~\cite{Tomino:2010xw}.

Finally, we compare our 3-Lie algebra model to the cubic supermatrix model of Smolin~\cite{Smolin:2000kc}, which has an additional $\mathfrak{osp}(1|32)$ symmetry algebra. This symmetry algebra was conjectured to be the appropriate one for M-theory. Using the Clifford algebra of $\FR^{1,10}$, we are indeed able to rewrite our 3-Lie algebra reduced model in an $\mathfrak{osp}(1|32)$-invariant fashion.

This paper is structured as follows: In section 2, we consider various deformations of the BLG theory and its dimensional reduction down to zero dimensions. We also describe the deformed IKKT model resulting from the recently proposed Higgs mechanism. Various stable solutions to our 3-Lie algebra reduced model are presented in section 3 and interpreted in terms of quantized Nambu-Poisson manifolds. In section 4, we attempt to make sense of noncommutative field theories on these quantized spaces. We conclude in section 5 with a comparison of our model to the cubic $\mathfrak{osp}(1|32)$-invariant supermatrix model of Smolin. An appendix at the end of the paper contains some details concerning 3-Lie algebras which are used in the main text.

\section{The 3-Lie algebra reduced model}

\subsection{Supersymmetric deformations of the BLG theory}

The BLG theory~\cite{Bagger:2007jr,Gustavsson:2007vu} is an $\CN=8$ supersymmetric Chern-Simons-matter theory in three dimensions with matter fields taking values in a metric 3-Lie algebra\footnote{See appendix A for the definitions and our notations related to 3-Lie algebras.} $\CA$ and a connection one-form taking values in the associated Lie algebra $\frg_\CA$. The matter fields consist of eight scalar fields $X^I$, $I=1,\ldots,8$ and their superpartners, which can be combined into a Majorana spinor $\Psi$ of $\sSO(1,10)$ satisfying $\Gamma_{012}\Psi=-\Psi$; throughout we denote $\Gamma_{M_1\cdots M_k}:= \frac1{k!}\, \Gamma_{[M_1}\cdots \Gamma_{M_k]}$. The Chern-Simons term is constructed using the alternative cyclic invariant form $\lbr-,-\rbr$ available on $\frg_\CA$ which is induced by the inner product $(-,-)$ on the 3-Lie algebra $\CA$. Altogether the action reads
\begin{equation}\label{eq:actionBLG}
\begin{aligned}
S_{\rm BLG}=\int\dd^3 x\ \Big(&-\tfrac{1}{2}\,\big(\nabla_\mu X^I,\nabla^\mu X^I\big)+\tfrac{\di}{2}\, \big(\bar{\Psi}, \Gamma^\mu\, \nabla_\mu \Psi\big)+\tfrac{\di}{4}\, \big(\bar{\Psi}, \Gamma_{IJ}[X^I,X^J,\Psi]\big)\\
&-\tfrac{1}{12}\, \big([X^I,X^J,X^K],[X^I,X^J,X^K]\big)+\tfrac{1}{2}\, \epsilon^{\mu \nu \lambda}\,\blbr A_\mu, \partial_\nu A_\lambda+\tfrac{1}{3}\, [A_\nu,A_\lambda] \brbr\Big)~,
\end{aligned}
\end{equation}
where $\mu,\nu,\lambda=0,1,2$ are indices for Euclidean coordinates on $\FR^{1,2}$. The matrices $\Gamma^\mu$, together with $\Gamma^I$, form the generators of the Clifford algebra $C\ell(\FR^{1,10})$. The covariant derivatives act according to
\begin{equation}
 \nabla_\mu X^I=\dpar_\mu X^I+A_\mu\, X^I:=\dpar_\mu X^I+A_\mu^{ab}\,D(\tau_a,\tau_b)X^I:=\dpar_\mu X^I+A_\mu^{ab}\,[\tau_a,\tau_b,X^I]~,
\end{equation}
where $\tau_a$ are generators of the 3-Lie algebra $\CA$. 

A well-known problem of this theory is that the only non-trivial 3-Lie algebra with positive definite invariant form $(-,-)$ is\footnote{See appendix A.} $A_4$. To circumvent this problem, Lorentzian 3-Lie algebras were introduced, but even this case is highly restrictive~\cite{DeMedeiros:2008zm}. Here, we will allow the matter fields to take values in the generalized 3-Lie algebras introduced in~\cite{Cherkis:2008qr}. As shown there, using generalized 3-Lie algebras will preserve at least four of the 16 supersymmetries of the original BLG theory. This brings us closer to the situation of $\sU(N)$ Yang-Mills theory, since these 3-Lie algebras allow for representations using matrices of arbitrary sizes as shown in~\cite{Cherkis:2008ha}. (An alternative direction would have been to work with the ABJM theory~\cite{Aharony:2008ug}, but this would involve working with complex 3-Lie algebras~\cite{Bagger:2008se}, which we want to avoid in our considerations.) 

The second deformation we introduce consists of adding mass and Myers-like flux terms given respectively by
\begin{equation}\label{eq:gendef}
\begin{aligned}
 S_{\rm mass}&=\int \dd^3 x\ \Big(-\tfrac{1}{2}\, \sum_{I=1}^8\, \mu^2_{1,I}\, \big(X^I,X^I\big)+\tfrac{\di}{2}\, \mu_2\, \big(\bar{\Psi},\Gamma_{3456}\Psi \big)\Big)~,\\[4pt]
 S_{\rm flux}&=\int \dd^3 x\ H^{IJKL}\, \big([X^I,X^J,X^K],X^L \big)~,
\end{aligned}
\end{equation}
where $H^{IJKL}$ is totally antisymmetric and can be thought of as originating from a four-form flux. A particularly interesting deformation is given by
\begin{equation}\label{eq:SUSYdef}
 \mu_{1,I}=\mu_2=\mu\eand H^{IJKL}=-\frac{\mu}{6}\, \left\{\begin{array}{cl}
                                                        \eps^{IJKL} & I,J,K,L\leq 4\\
							\eps^{(I-4)(J-4)(K-4)(L-4)} & I,J,K,L\geq 5 \\
0 & \mbox{otherwise}
                                                       \end{array}\right.~.
\end{equation}
This deformation was studied first in~\cite{Gomis:2008cv}, see also~\cite{SheikhJabbari:2004ik,Hosomichi:2008qk}. It is closely related to the deformation giving rise to the BMN matrix model~\cite{Berenstein:2002jq} and homogeneous gravitational wave backgrounds, as we will discuss later on. It explicitly breaks the R-symmetry group $\sSO(8)$ down to $\sSO(4)\times \sSO(4)$, but preserves all 16 supersymmetries if the matter fields live in a 3-Lie algebra. If the fields take values in a generalized 3-Lie algebra, then at least four supersymmetries will be preserved.

The third deformation we admit is the addition of a Yang-Mills term
\begin{equation}
 \begin{aligned}
  S_{\rm YM}=\frac{1}{4\gamma^2}\, \int \dd^3 x\ \blbr F_{\mu\nu},F^{\mu\nu}\brbr
 \end{aligned}
\end{equation}
 to the action. In three dimensions the Yang-Mills action is an
 irrelevant term in the quantum field theory. In the infrared the
 renormalization group flow will cause this term to vanish, and
 theories with different values of the Yang-Mills coupling $\gamma$
 become indistinguishable. We therefore decide to allow this term in our action.

\subsection{Dimensional reduction of the deformed BLG theory}

The dimensional reduction of the theory defined by the action $S=S_{\rm BLG}+S_{\rm mass}+S_{\rm flux}+S_{\rm YM}$ is now straightforward. We reduce the covariant derivatives $\nabla_\mu$ to an action of the gauge potential $A_\mu$, which yields
\begin{equation}\label{eq:action3M}
\begin{aligned}
 S=&-\tfrac{1}{2}\,\big(A_\mu X^I,A^\mu X^I\big)+\tfrac{\di}{2}\, \big(\bar{\Psi}, \Gamma^\mu\, A_\mu \Psi\big)\\&
 -\tfrac{1}{2}\, \sum_{I=1}^8\, \mu^2_{1,I}\, \big(X^I,X^I\big)+\tfrac{\di}{2}\, \mu_2\, \big(\bar{\Psi},\Gamma_{3456}\Psi \big)
 +H^{IJKL}\,\big([X^I,X^J,X^K],X^L\big)\\
&+\tfrac{\di}{4}\,\big(\bar{\Psi}, \Gamma_{IJ}[X^I,X^J,\Psi]\big)-\tfrac{1}{12}\,\big([X^I,X^J,X^K],[X^I,X^J,X^K]\big)\\
&+\tfrac{1}{6}\, \epsilon^{\mu \nu \lambda}\, \blbr A_\mu,[A_\nu,A_\lambda]\brbr+\frac{1}{4\gamma^2}\, \blbr[A_\mu,A_\nu],[A^\mu,A^\nu]\brbr~.
\end{aligned}
\end{equation}
This model has the same amount of supersymmetry as the original unreduced field theory. However, it is only invariant under the group $\sSO(1,2)\times \sSO(8)$ instead of the desired 11-dimensional Lorentz group $\sSO(1,10)$, which is due to the dichotomy of gauge and matter fields in the original BLG theory. This is in marked contrast to the IKKT matrix model which arises from dimensional reduction of maximally supersymmetric Yang-Mills theory to zero dimensions, and therefore exhibits manifest $\sSO(1,9)$ invariance.

Nevertheless, we still consider the model \eqref{eq:action3M} to be
interesting for the following reasons. First of all, we will show
below that in a certain limit we can reduce it to the IKKT matrix
model and therefore at least restore $\sSO(1,9)$ invariance in this
limit. Second, the alternative model based on $\sSO(1,10)$-invariant
constructions involving 3-Lie algebras breaks too many of the
supersymmetries~\cite{Furuuchi:2009ax}. And third, almost all the
solutions we will be interested in will solely rely on the pure matter
part of the action, in which our model agrees with the
$\sSO(1,10)$-invariant model of~\cite{Furuuchi:2009ax} (see also~\cite{Sato:2009mf}).

\subsection{Reduction to the IKKT matrix model}

If one assumes that the BLG theory describes M2-branes, then one should be able to reduce the BLG theory to the effective description of D2-branes which is given by maximally supersymmetric Yang-Mills theory in three dimensions. In the paper~\cite{Mukhi:2008ux}, Mukhi and Papageorgakis proposed such a reduction procedure for the BLG theory with 3-Lie algebra $\CA=A_4$, which reduces to $\CN=8$ supersymmetric Yang-Mills theory with gauge group $\sSU(2)$. Below we briefly review this reduction by going through the corresponding procedure for the dimensionally reduced model.

We start from our model \eqref{eq:action3M} with 3-Lie algebra $A_4$, whose generators are denoted $e_i$, $i=1,2,3,4$, and assume that one of the scalar fields, corresponding to the M-theory direction, develops a vacuum expectation value (vev) which is proportional to the radius $R$ of the M-theory circle. Using the $\sSO(4)$-invariance of $A_4$, we can align this vev in the $e_4$ direction so that
\begin{equation}
 \langle X^8 \rangle=-\frac{R}{\ell_p^{3/2}}\, e_4=-g_{\rm YM}\, e_4~,
\end{equation}
where $\ell_p$ and $g_{\rm YM}$ are the 11-dimensional Planck length and the Yang-Mills coupling constant, respectively. We now expand the action \eqref{eq:action3M} around this vev by rewriting
\begin{equation}\label{eq:X8vev}
 X^8=-g_{\rm YM}\, e_4+Y^8~,
\end{equation}
where $Y^8\in\CA$ still has components along the $e_4$ direction. The 3-brackets containing $X^8$ reduce according to
\begin{equation}\label{eq:reduction}
 [A,B,X^8]=g_{\rm YM}\, [A,B,e_4]+[A,B,Y^8]~,~~~A,B\in\CA~,
\end{equation}
and in the strong coupling limit, i.e.\ for large values of $g_{\rm YM}$, 3-brackets containing $X^8$ reduce to the Lie bracket of $\aso(3)$ due to $[e_i,e_j,-e_4]=\eps_{ijk4}\, e_k$. It is easy to see that the potential terms in \eqref{eq:action3M} containing matter fields reduce to the corresponding terms of the IKKT matrix model for $\gamma\rightarrow \infty$ and $\mu=0$. 

The reduction of the terms involving the gauge potential is slightly more involved. One considers the splitting $\frg_{A_4}=\aso(4)\cong \aso(3)\oplus\aso(3)$ according to
\begin{equation}\label{eq:splitting}
 A_\mu=A_\mu^{ij}\, D(e_i,e_j)=A_\mu^i\, D(e_i,e_4)+B_\mu^i\, \tfrac{1}{2}\, \eps_{ijk}\, D(e_j,e_k)~.
\end{equation}
In the action \eqref{eq:action3M}, the field $B_\mu^i$ appears in the strong coupling limit only algebraically, and its equation of motion reads
\begin{equation}\label{eq:eom}
 B_\mu^i=\frac{1}{2g_{\rm YM}}\, \eta_{\mu\nu}\,
 \eps^{\nu\rho\lambda}\, \eps^{ijk}\, A^j_\rho\,
 A^k_\lambda-\frac{1}{2g_{\rm YM}}\, \eps^{ijk}\, A_\mu^j\, X^{8\, k}~,
\end{equation}
where $\eta_{\mu\nu}$ denotes the Minkowski metric on $\FR^{1,2}$. The reduction \eqref{eq:reduction} together with the splitting \eqref{eq:splitting} and the equation of motion \eqref{eq:eom} reduce the action \eqref{eq:action3M} with $\gamma\rightarrow \infty$ and $\mu=0$ to the action of the IKKT matrix model with gauge group $\asu(2)\cong\aso(3)$,
\begin{equation}
\begin{aligned}\label{eq:actionIKKT}
 S_{\rm IKKT}=-\tfrac{1}{4}\,\big([\CCX_M,\CCX_N],
 [\CCX^M,\CCX^N]\big)+\tfrac{\di}{2}\, \big( \bar{\Psi}, \Gamma^M[\CCX_M,\Psi]\big)~.
\end{aligned}
\end{equation}
Here we combined the fields $(A_\mu,X^I)$ with $\mu=0,1,2$ and
$I=1,\ldots,7$ into $\CCX^M$ with $M=0,1,\ldots,9$, and absorbed the
coupling $g_{\rm YM}$ into a rescaling of fields. The invariant
bilinear inner product in this instance coincides with the Cartan-Killing form on the Lie algebra $\asu(2)$, $(\CCX,\CCY)=\tr(\CCX\, \CCY)$. This matrix model possesses $32$ supersymmetries.

In the strong coupling limit, the Myers-like term in \eqref{eq:action3M} is reduced according to
\begin{equation}
\begin{aligned}
 H^{IJKL}\, \big([X^I,X^J,X^K],X^L \big)~~\longrightarrow~~4g_{\rm
   YM}\, H^{IJK8}\, \big([X^I,X^J], X^K \big)~,
\end{aligned}
\end{equation}
and this is the Myers term appearing in the deformation of the BFSS
model to the BMN matrix model~\cite{Berenstein:2002jq}. Including the
mass terms, the deformation terms reduce to
\begin{equation}
\begin{aligned}\label{eq:gendefIKKT}
 S_{\rm mass+flux}=& 
-\tfrac{1}{2}\, \sum_{I=1}^7\, \mu_{1,I}^2\,\big(
\CCX^{I+2}, \CCX^{I+2}\big) +\tfrac{\di}{2}\, \mu_2\, \big(\bar{\Psi}, \Gamma_{3456}\Psi \big)
\\&+4g_{\rm YM}\, \sum_{I,J,K=1}^7\, H^{IJK8}\, \big(
[\CCX^{I+2},\CCX^{J+2}], \CCX^{K+2}\big)~.
\end{aligned}
\end{equation}
If the 3-Lie algebra $\CA$ is not $A_4$, a vev for one of the fields
will still reduce the bosonic part of the potential to an ordinary Lie
algebra expression in the strong coupling limit. The reduction of the gauge part of the action, however, will break down in general.

\section{Classical solutions}

\subsection{Equations of motion}

The classical equations of motion of our 3-Lie algebra model \eqref{eq:action3M} with a metric 3-Lie algebra $\CA$ read
\begin{equation}\label{eq:eoms3M}
 \begin{aligned}
A_\mu\, A^\mu\, X^I-\mu_{1,I}^2\, X^I-\di\, [\bar{\Psi},X^J,\Gamma_{IJ} \Psi]\hspace{6cm}&\\+\, \tfrac{1}{2}\, \big[X^J,X^K,[X^J,X^K,X^I]\big]+4H^{IJKL}\, [X^J,X^K,X^L] \ =\ &0~,\\[4pt]
\Gamma^\mu\, A_\mu \Psi+\mu_2\, \Gamma_{3456}\Psi+\tfrac{1}{2}\, \Gamma_{IJ}[X^I,X^J,\Psi] \ = \ &0~,\\[4pt]
 \tfrac{1}{2}\,\epsilon^{\mu\nu\lambda}\, [A_\nu, A_\lambda]-\tfrac{1}{\gamma^2}\, \big[A_\nu,[A^\nu,A^\mu]\big]-D(X^I,A^\mu X^I)+\tfrac{\di}{2}\, D(\bar{\Psi},\Gamma^\mu \Psi) \ = \ &0~.
 \end{aligned}
\end{equation}
The classical equations of motion of the IKKT matrix model \eqref{eq:actionIKKT}, i.e.\ the strong coupling limit of the 3-Lie algebra model \eqref{eq:action3M}, read
\begin{equation}
\begin{aligned}
{}\big[\CCX_N,[\CCX^N,\CCX^M]\big]-\tfrac{\di}{2}\,
\Gamma^M_{\alpha\beta}\, \{\Psi^\beta,\bar{\Psi}^\alpha\}+\Delta^M \ = \ &0~,\\[4pt]
\Gamma^M\, [\CCX_M,\Psi]+\mu_2\, \Gamma_{3456}\Psi \ = \ &0~, 
\end{aligned}
\label{eq:IKKTeom}\end{equation}
where $\alpha,\beta$ are spinor indices of a Majorana-Weyl spinor of $\sSO(1,9)$ and the deformation contribution is
\begin{equation}
 \Delta^M=\left\{
\begin{array}{ll}
-\mu^2_{1,M-2}\, \CCX^M+12 g_{\rm YM}\, \displaystyle{\sum\limits_{I,J=1}^7}\, H^{IJ(M-2)8}\, [\CCX^{I+2},\CCX^{J+2}] & \mbox{for} \quad3\leq M \leq 9\\
0 & \mbox{for} \quad M=0,1,2
\end{array}
\right.~.
\end{equation}

In the following we will study solutions to these equations and examine their classical stability. Recall that in the IKKT model, one usually starts with gauge group $\sU(N)$ for $N$ ``large enough'' and then considers solutions which correspond to the branching of $\sU(N)$ to some other Lie group. For example, for the fuzzy sphere solutions arising in the IKKT model deformed by a Chern-Simons term, one studies branchings $\sU(N)\rightarrow \sSU(2)$. There is no direct analogue of the ``universal gauge symmetry'' $\sU(N)$ for 3-Lie algebras; in particular there is no family of 3-Lie algebras with positive definite metric except for direct sums of $A_4$~\cite{Nagy:2007aa,figueroaofarrill-2004-49}. We can switch to generalized 3-Lie algebras (for which the existence of continuous families follows from the representations found in~\cite{Cherkis:2008ha}), and assume that the generalized 3-Lie algebra we started from is ``large enough'' to contain all our solutions. Note that the equations of motion for generalized 3-Lie algebras would be slightly different from \eqref{eq:eoms3M}. However, we want to find solutions which can be interpreted as quantized Nambu-Poisson manifolds in the sense of~\cite{DeBellis:2010pf}, and we will restrict ourselves to solutions which form 3-Lie algebras and therefore satisfy \eqref{eq:eoms3M}.

\subsection{Fuzzy spheres}

As it is the most prominent 3-Lie algebra, let us start with a solution involving $A_4$. For this, we choose the supersymmetric deformation \eqref{eq:SUSYdef} to obtain a natural $\sSO(4)$ symmetry group, which matches the associated Lie group of $A_4$. We put $A_\mu=\Psi=0$. As scalar fields, we choose
\begin{equation}\label{eq:solS3}
 X^i=\alpha\,  e_i~,~~~X^{i+4}=0~,~~~\mbox{with}~~~\alpha^4+\tfrac{4}{3}\, \mu\, \alpha^2+\tfrac{1}{3}\, \mu^2=0~,
\end{equation}
where $ e_i$, $i=1,2,3,4$ are generators of $A_4$. This solution corresponds to a fuzzy three-sphere~\cite{Guralnik:2000pb} in the sense of~\cite{DeBellis:2010pf} with a radius proportional to $\sqrt\mu$. The relation between $A_4$ and fuzzy $S^3$ has been pointed out many times starting with~\cite{Bagger:2006sk}. The first derivation of fuzzy three-spheres from the BLG theory by including flux deformations was given in~\cite{Gomis:2008cv}.
 
We can now compute the Hessian of the action $\frac{\delta^2 S}{\delta
  X^{i\, a}\, \delta X^{j\, b}}$, where $\delta X^{i \, a}$ describes
the variation of $X^i$ in the 3-Lie algebra direction $ e_a$. One
finds a $16\times 16$ matrix with eigenvalues $(0,2,6)\, \mu^2$ occuring in multiplicities $(6,9,1)$. The six flat directions correspond to variations rotating the fuzzy sphere. (The other eigenvalues correspond to ``squashing'' the fuzzy sphere in various ways.) We conclude that the solution \eqref{eq:solS3} is indeed a stable stationary point of the action \eqref{eq:action3M}. Moreover, like the ground states used in~\cite{Gomis:2008cv}, our solutions are invariant under the full set of 16 supersymmetries of the deformed action. This can be checked explicitly by noting that the supersymmetry transformation for $A_\mu=0$ reads~\cite{Gomis:2008cv}
\begin{equation}
\delta_\eps X^I=\di\, \bar\eps\, \Gamma^I\Psi \ , \qquad
 \delta_\eps \Psi=-\tfrac{1}{6}\,[X^I,X^J,X^K]\,\Gamma^{IJK}\eps-\mu\, \Gamma_{3456}\, \Gamma^I\, X^I\eps~,
\label{eq:SUSYeps}\end{equation}
where $\eps$ is a constant Majorana spinor of $\sSO(1,10)$ satisfying $\Gamma_{012}\eps=\eps$, and hence our fuzzy three-sphere background satisfies the supersymmetry condition $\delta_\eps X^I=0=\delta_\eps\Psi$.

We can now apply the Higgs mechanism. We assume that one of the scalar fields acquires a vev and perform a strong coupling expansion. Let us choose $X^4=g_{\rm YM}\,  e_4+Y^4$ and take a double scaling limit $g_{\rm YM},\mu\rightarrow \infty$ with $\hat\mu= \frac{\mu}{g_ {\rm YM}}$ fixed. The equations of motion reduce to
\begin{equation}
\begin{aligned}
 {}\big[X^j,[X^j,X^i]\big]-2\hat\mu\, \eps^{ijk}\,[X^j,X^k]&=0~,\\[4pt]
 \big[X^j,X^k,[X^j,X^k,X^4]\big]+2\hat\mu\, \eps^{4jkl}\, [X^j,X^k,X^l]&=0~,
\end{aligned}
\end{equation}
for $i=1,2,3$. The first equation is the equation of motion of the IKKT model with a Myers term and its solution is a fuzzy two-sphere, i.e.\ the matrices $X^i$ take values in $\asu(2)$. The second equation requires the Lie algebra $\asu(2)$ to be consistently embedded in $A_4$. Altogether, we see that the fuzzy two-sphere originates as the strong coupling limit of the fuzzy three-sphere. Geometrically, we reduced the fuzzy three-sphere to its equator with radius $g_{\rm YM}$, which corresponds to the fuzzy two-sphere solution. This is {\em not} the projection of the Hopf fibration $S^1\hookrightarrow S^3\rightarrow S^2$.

Note that our deformation is very similar to that of the BMN model,
which can be considered as the BFSS model on a non-trivial Hpp-wave
background. The fuzzy two-sphere solution is in that case interpreted
as giant gravitons, i.e.\ M2-branes wrapping the fuzzy $S^2$ with
certain kinematical properties. The supersymmetric deformation
\eqref{eq:SUSYdef} has been holographically linked
in~\cite{Gomis:2008cv} to the matrix model description of the
maximally supersymmetric type IIB plane wave in discrete light-cone
quantization; this Hpp-wave background is a ten-dimensional
Cahen-Wallach symmetric space which arises as a Penrose limit of the
near horizon black hole geometry ${\rm AdS}_5\times S^5$ in type~IIB
supergravity~\cite{Blau:2002mw}. It has metric
\begin{equation}
\dd s^2=2\,\dd x^+\, \dd x^- +\sum_{I}\, \Big(\dd x_I^2- \mbox{$\frac14$}\, \mu^2\,x_I^2 \, (\dd x^+)^2\Big) \ ,
\label{eq:Hppwave}\end{equation}
and constant null self-dual Ramond-Ramond five-form flux $H_{\rm RR}= \mu\, \dd x^+\wedge(\dd x^{1234}+\dd x^{5678})$, where the sum runs over $I=1,\dots,8$ and $\dd x^{IJKL}:=\dd x^I\wedge\dd x^J\wedge\dd x^K\wedge \dd x^L$. The fuzzy three-sphere solution obtained here was identified in~\cite{Gomis:2008cv} with longitudinal D3-brane giant gravitons in this background.

\subsection[R**3-lambda and the noncommutative plane]{$\FR^3_\lambda$ and the noncommutative plane}

In the (undeformed) IKKT matrix model, the simplest classical solution is given by operators $\CCX^1=\lambda_1$ and $\CCX^2=\lambda_2$, where $\lambda_1$ and $\lambda_2$ are the generators of the Heisenberg algebra $[\lambda_1,\lambda_2]=\theta\,\unit$, $\theta\in\FR$. The D-brane interpretation of this solution involves D$(-1)$-branes described by the scalar fields in a background $B$-field proportional to $\theta^{-1}$ which are smeared out into a D1-brane, whose worldvolume is the noncommutative space $\FR^2_\theta$. This solution can be evidently extended to direct sums of $\FR^2_\theta$, by demanding that further pairs of scalar fields satisfy the Heisenberg algebra. Note, however, that there is an issue with the normalizability of the central element $\unit$, as the Heisenberg algebra only has infinite-dimensional unitary representations.

The classical vacuum state of the reduced model with action (\ref{eq:actionIKKT}) is given by commuting matrices $\CCX^M$. Noncommutative spacetime arises instead as a vacuum configuration of the \emph{twisted} reduced model with action
\begin{equation}
\begin{aligned}\label{eq:actionIKKTtwisted}
 \widetilde{S}_{\rm IKKT}=-\tfrac{1}{4}\,\big([\CCX_M,\CCX_N] -\theta_{MN}\,\unit ,
 [\CCX^M,\CCX^N]-\theta^{MN}\, \unit \big)+\tfrac{\di}{2}\, \big( \bar{\Psi}, \Gamma^M[\CCX_M,\Psi]\big)~,
\end{aligned}
\end{equation}
where the ``twist'' $\theta_{MN}$ is generically a $10\times10$ constant antisymmetric real matrix; in the special case considered above only $\theta_{12}=\theta$ is nonzero. The solutions with $\CCX^M=\lambda_M$, $[\lambda_M,\lambda_N]=\theta_{MN}\,\unit$ correspond to BPS-saturated backgrounds which preserve half the $32$ supersymmetries. Upon introducing the covariant coordinates
\begin{equation}\label{eq:covcoords}
\CCX^M=\lambda_M+\theta_{MN}\, \CCA^N \ ,
\end{equation}
corresponding to expansion around the infinitely-extended D-branes in the original IKKT model, one obtains the action for $\sU(1)$ noncommutative supersymmetric Yang-Mills theory with 16 supercharges~\cite{Aoki:1999vr} and trivial vacuum state $\CCA^M=0$; the gauge fields $\CCA^M$ are interpreted as dynamical fluctuations about the noncommutative spacetime. To obtain the action for noncommutative Yang-Mills theory with $\sU(m)$ gauge group, corresponding to the background of $m$ coincident D-branes, one expands around the vacuum $\CCX^M=\lambda_M \otimes\unit_m$. We will return to these expansions later on.

Exactly the same sort of configurations arise in our model. The configuration $X^i=\tau_i$ for $i=1,2,3$ and $X^I=0$ for $I=4,5,6,7,8$, where $\tau_1,\tau_2,\tau_3,\unit$ are generators of the Nambu-Heisenberg 3-Lie algebra $\CA_{\rm NH}$,
\begin{equation}
 [\tau_1,\tau_2,\tau_3]=\theta\, \unit~,~~~~[\unit,\tau_i,\tau_j]=0~,
\end{equation}
forms evidently a solution to our equations of motion \eqref{eq:eoms3M} in the absence of fluxes and for $A_\mu=\Psi=0$. This 3-Lie algebra was originally considered by Nambu~\cite{Nambu:1973qe} in the context of generalizations of Hamiltonian dynamics and their quantizations. Recently it was derived as a boundary condition on the geometry of an M5-brane in the M2--M5 brane system in a constant background $C$-field~\cite{Chu:2009iv}. It has associated Lie algebra $\frg_{\CA_{\rm NH}}\cong\mathbb{R}^6$.

The solution $X^I=\tau_I$, $[\tau_I,\tau_J,\tau_K]=\Theta_{IJK}\, \unit$, with $\Theta^{IJK}$ a constant real three-form flux, describes the vacuum state of the ``twisted'' version of the scalar potential of the action (\ref{eq:action3M}) based on the 3-Lie algebra $\CA=\CA_{\rm NH}$ in the absence of masses and fluxes, which generically reads
\begin{equation}
\widetilde{V}(X)= -\tfrac{1}{12}\,\big([X^I,X^J,X^K]-\Theta^{IJK}\, \unit,[X^I,X^J,X^K] -\Theta^{IJK}\,\unit\big) \ .
\label{eq:twistedBLGpot}\end{equation}
In fact, this solution preserves 16 supersymmetries. This follows from the general fact that the model (\ref{eq:action3M}) based on a 3-Lie algebra $\CA$ with central element $\unit$ for the configuration (\ref{eq:SUSYdef}) possesses an additional 16 kinematical supersymmetries~\cite{Gomis:2008cv}
\begin{equation}
\tilde\delta_\xi X^I=0 \ , \qquad \tilde\delta_\xi\Psi=\xi\, \unit \ ,
\label{eq:SUSYxi}\end{equation}
where $\xi$ is a constant spinor of $\sSO(1,10)$ satisfying $\Gamma_{012}\xi=-\xi$. Setting $X^I=\tau_I$, $\mu=0$ and $\xi=\frac16\,\Theta_{IJK}\,\Gamma^{IJK}\eps$ in the supersymmetry transformations (\ref{eq:SUSYeps}) and (\ref{eq:SUSYxi}), one finds the relations
\begin{equation}
(\delta_\eps+\tilde\delta_\xi)X^I=0 \ , \qquad (\delta_\eps+\tilde\delta_\xi)\Psi=0 \ ,
\end{equation}
and hence half of the 32 supersymmetries are preserved in these backgrounds. This is consistent with the calculation of~\cite{Tomino:2010xw} which shows that the one-loop vacuum energy of these backgrounds vanishes.

An interpretation of the Nambu-Heisenberg algebra in terms of quantized Nambu-Poisson manifolds is given by the noncommutative space $\FR^3_\lambda$~\cite{DeBellis:2010pf}. If we assume that $X^3$ acquires a vev proportional to a coupling constant, then in the strong coupling limit the Nambu-Heisenberg algebra reduces to the ordinary Heisenberg algebra. In this sense, the noncommutative plane $\FR^2_\theta$ can be regarded as the strong coupling limit of $\FR^3_\lambda$. Again, we can extend our solution to the direct sum $\FR^3_\lambda\oplus\FR^3_\lambda$ by demanding that three more of the scalar fields form another copy of the Nambu-Heisenberg 3-Lie algebra; this is the quantized geometry relevant to an M5-brane in a constant $C$-field background~\cite{Chu:2009iv,DeBellis:2010pf}. As in the case of the IKKT matrix model, there is a problem with the normalizability of the 3-central element $\unit$; the compatibility condition (cf.\ appendix A) forbids us to assign finite norm to $\unit$. 
There is a natural extension of the Heisenberg Lie algebra and the Nambu-Heisenberg 3-Lie algebra which avoids the normalizability problem; we describe these extensions below.

\subsection{Homogeneous plane wave backgrounds}

The homogeneous plane wave with metric (\ref{eq:Hppwave}), and supported by a Neveu-Schwarz flux, can be constructed as the group manifold of the twisted Heisenberg group whose Lie algebra is an extension of the Heisenberg algebra by one additional generator $J$ defined by
\begin{equation}\label{eq:NWalg}
[\lambda_M,\lambda_N]=\theta_{MN}\,\unit \ , \qquad [J,\lambda_M]=\theta_{MN}\,\lambda_N \ , \qquad [\unit,\lambda_M]=[\unit,J]=0 \ .
\end{equation}
The simplest case is $\theta_{MN}=\eps_{MN}$, $M,N=1,2$ corresponding to the Nappi-Witten algebra~\cite{Nappi:1993ie}, which is a non-semisimple Lorentzian Lie algebra of dimension four. The Lie brackets (\ref{eq:NWalg}) are then those of the universal central extension of $\mathfrak{iso}(2)$.

Let us now consider the mass and flux deformations of the IKKT model \eqref{eq:gendefIKKT} given by
\begin{equation}
 \mu_{1,6}=\mu_{1,7}=\mu~,~~~~H^{5678}=h~,
\end{equation}
where all other mass terms and components of $H$ vanish. We choose the ansatz
\begin{equation}\label{eq:ppAnsatzIKKT}
 \CCX^6=\alpha\, \unit~,~~~~\CCX^7=\beta\, J~,~~~~\CCX^8=\gamma\, \lambda_1~,~~~~\CCX^9=\gamma\, \lambda_2 \ ,
\end{equation}
with $\CCX^M=0=\Psi$ for $M=0,1,2,3,4,5$, for our solution. Then the equations of motion (\ref{eq:IKKTeom}) are satisfied if 
\begin{equation}
 \mu^2=(24g_{\rm YM}\, h)^2\eand \beta=-24g_{\rm YM}\, h~,
\end{equation}
while the parameters $\alpha$ and $\gamma$ are arbitrary. These solutions are not supersymmetric.

This noncommutative background can be regarded as a linear Poisson
structure on a four-dimensional Hpp-wave. The invariant,
non-degenerate symmetric bilinear forms on the Nappi-Witten Lie algebra
are parametrized by a real number $b$ and are defined by
\begin{equation}\label{eq:NWform}
 (\lambda_i,\lambda_j)=\delta_{ij}~,~~~~ (\unit,J)=1~,~~~~(J,J)=b
\end{equation}
for $i,j=1,2$, with all other pairings vanishing.
Then the group manifold possesses a homogeneous bi-invariant Lorentzian metric defined by the pairing of the left-invariant Cartan-Maurer one-forms as
\begin{equation}
\dd s_4^2=(g^{-1}\, \dd g,g^{-1}\, \dd g) \ .
\end{equation}
We can parametrize group elements $g$ as
\begin{equation}
g=\exp\big({\rm e}^{\di\, \beta\, x^+/2}\, z\, \CCZ_++{\rm e}^{-\di\,\beta\, x^+/2}\, \overline{z}\, \CCZ_- \big)\, \exp\big(x^-\, \CCX^6+ x^+\, \CCX^7\big) \ ,
\end{equation}
where $\CCZ_\pm=\CCX^8\pm\di\, \CCX^9$, $x^\pm\in\FR$ and $z\in\FC$. Then the metric in these global coordinates reads
\begin{equation}
\dd s_4^2=2\alpha\,\beta\, \dd x^+\, \dd x^-+\gamma^2\, |\dd z|^2 -\tfrac{1}{4}\,\beta^2\, \big(\gamma^2\, |z|^2 -b\big) \, (\dd x^+)^2 \ ,
\label{eq:4Dppwave}\end{equation}
which is the standard form of the plane wave metric of a four-dimensional Cahen-Wallach symmetric spacetime in Brinkman coordinates. This spacetime is further supported by a constant null Neveu-Schwarz three-form flux
\begin{equation}
H_{\rm NS}=-\tfrac{1}{3}\,\big(g^{-1}\, \dd g,\dd(g^{-1}\,\dd g)\big) = 2\,\di\, \beta\, \gamma^2\, \dd x^+\wedge \dd z\wedge \dd\overline{z} \ ,
\end{equation}
which is proportional to the flux deformation $h$ of the matrix model. 

The Hessian for this solution is a $16\times 16$ matrix with
eigenvalues $(0,1,2,3)\,\mu^2$ of multiplicities $(6,1,8,1)$. The six flat directions correspond to the following symmetries of the matrix model defined by (\ref{eq:actionIKKT}) and (\ref{eq:gendefIKKT}) with the appropriate inner product (\ref{eq:NWform}). One direction corresponds to the $\sU(1)$ subgroup of the plane wave isometry group rotating the transverse space $z\in\FC$. Three directions correspond to translations of the Nappi-Witten generators by multiples of the central element $\unit$. Of these, only shifts of the generator $J$ are inner automorphisms of the Lie algebra (\ref{eq:NWalg}); in particular, the automorphism $J\mapsto J-b\,\unit$ can be used to set the parameter $b$ to $0$ in (\ref{eq:NWform}), which is equivalent to the redefinition $x^-\to x^--\frac18\, \frac{\gamma^2\,\beta}\alpha\,b\, x^+$ in the plane wave metric (\ref{eq:4Dppwave}). The shifts in $\lambda_i$ are isometries which translate the transverse space along the null direction $x^+$. Another direction corresponds to scale transformations $\unit\to{\rm e}^\zeta\, \unit$, which becomes a Lie algebra automorphism after redefining $\lambda_i\to\de^{\zeta/2}\, \lambda_i$. The final symmetry of the action corresponds to the simultaneous scale transformations $J\to\de^{-\zeta}\, J$, $\lambda_i\to\de^\zeta\, \lambda_i$.

This Hpp-wave background is thus a stable solution of the deformed IKKT matrix model. It arises as a Penrose limit of the maximally supersymmetric black hole solution with near horizon geometry ${\rm AdS}_2\times S^2$ in four-dimensional toroidal compactification of string theory and M-theory, or alternatively of the near horizon region of NS5-branes~\cite{Blau:2002mw}. Extending this solution by an additional noncommutative plane gives a Cahen-Wallach space which is a Penrose limit of the near horizon geometry ${\rm AdS}_3\times S^3$ of the self-dual string in six dimensions~\cite{Blau:2002mw}. Field theory on this noncommutative background has been formulated and described in~\cite{Halliday:2006qc}. Solutions of the IKKT model corresponding to gravitational plane waves have also been found in~\cite{Rivelles:2002ez}.

There is an analogous extension of the Nambu-Heisenberg 3-Lie algebra given by
\begin{equation}
[\tau_I,\tau_J,\tau_K]=\Theta_{IJK}\,\unit~,~~~~[J,\tau_I,\tau_J]=\Theta_{IJK}\,
  \tau_K~,~~~~[\unit,\tau_I,\tau_J]= [\unit,\tau_I,J]=0 \ .
\end{equation}
Again we focus on the simplest case $\Theta_{IJK}=\eps_{IJK}$, $I,J,K=1,2,3$. This is the {\em Nappi-Witten 3-Lie algebra} $\CA_{\rm NW}$ which is the semisimple indecomposable Lorentzian 3-Lie algebra obtained by double extension from the Lie algebra $\mathfrak{so}(3)$~\cite{DeMedeiros:2008zm}. Its associated Lie algebra is $\mathfrak{g}_{\mathcal{A}_{\rm NW}}\cong\mathfrak{iso}(3)$. In contrast to the Nambu-Heisenberg 3-Lie algebra, we can turn $\CA_{\rm NW}$ into a metric 3-Lie algebra by defining the symmetric bilinear form
\begin{equation}
 (\tau_i,\tau_j)=\delta_{ij}~,~~~~(\unit,J)=-1 ~,~~~~(J,J)=b
\end{equation}
for $i,j=1,2,3$ and arbitrary $b\in\FR$, with all other pairings equal to $0$.

We can now find a similar solution to our 3-Lie algebra model, if we choose the background \eqref{eq:gendef} with mass and flux terms
\begin{equation}
 \mu_{1,6}=\mu_{1,7}=\mu_{1,8}=\mu~,~~~~H^{5678}=h~,
\end{equation}
and all other mass terms and components of $H$ are zero. The obvious generalization of the ansatz \eqref{eq:ppAnsatzIKKT} to the 3-Lie algebra model reads
\begin{equation}
 X^4=\alpha\, \unit~,~~~~X^5=\beta\, J~,~~~~X^6=\gamma\, \tau_1~,~~~~X^7=\gamma\, \tau_2~,~~~~X^8=\gamma\, \tau_3~,
\end{equation}
with $A_\mu=0=\Psi$ and $X^I=0$ for $I=1,2,3$, and from the equations of motion we obtain conditions on the parameters
\begin{equation}
 \mu^2=(8h)^2~,~~~~\beta=-\frac{8h}{\gamma}~,
\end{equation}
while the parameters $\alpha$ and $\gamma$ are again arbitrary. It is natural to associate this solution with the extension of the pp-wave geometry (\ref{eq:4Dppwave}) by an additional transverse direction $y\in\FR$,
\begin{equation}
\dd s_5^2=2\alpha\,\beta\, \dd x^+\, \dd x^-+\gamma^2\, \big(|\dd z|^2+\dd y^2 \big) -\tfrac{1}{4}\,\beta^2\, \big(\gamma^2\, (|z|^2+y^2)-b \big) \, (\dd x^+)^2 \ .
\end{equation}
This five-dimensional Cahen-Wallach space arises as a Penrose limit of an ${\rm AdS}_2\times S^3$ background, which corresponds to the near horizon geometry of black hole solutions for $\CN=2$ supergravity in five dimensions~\cite{Blau:2002mw}. 

The Hessian of this solution is a $25\times25$ matrix with eigenvalues $(0,1,2,3,4,5)\,\mu^2$ of multiplicity $(8,3,3,5,3,3)$. Again the eight flat directions correspond to the $\sSO(3)$ subgroup of the plane wave isometry group generating rotations of the transverse space $(z,y)\in\FC\times\FR\cong \FR^3$, to null translations of the transverse space, to automorphisms $J\mapsto J-b\,\unit$ of the Nappi-Witten 3-Lie algebra, and to conformal rescalings of the 3-central element $\unit$. This background is thus a stable solution of the 3-Lie algebra reduced model (\ref{eq:action3M}).

\subsection{Fuzzy hyperboloids}

As a side remark, we note that for finite $\gamma$ the pure gauge part of the action (\ref{eq:action3M}) corresponds to the IKKT model deformed by a Myers term. Turning off the matter fields $X^I=\Psi=0$, the equations of motion read
\begin{equation}
  \tfrac{1}{2}\, \epsilon^{\mu\nu\lambda}\, [A_\nu, A_\lambda]-\tfrac{1}{\gamma^2}\, \big[A_\nu,[A^\nu,A^\mu]\big]=0~.
\end{equation}
A solution to these equations is given by $A_\mu=-\frac{\gamma^2}{2}\, \sigma_\mu$, where $\sigma_\mu$ generate the $\aso(2,1)$ Lie algebra $[\sigma_\mu,\sigma_\nu]=\eps_{\mu\nu\kappa}\, \eta^{\kappa\lambda}\, \sigma_\lambda$. This background can be regarded as coordinates on a fuzzy hyperboloid.\footnote{In fact, it corresponds to the one-point compactification of this hyperboloid; see the discussion in~\cite{DeBellis:2010pf}.}

If we had performed a Wick rotation of the action~\eqref{eq:action3M},
then we would have obtained solutions $A_\mu=-\frac{\gamma^2}{2}\,
\sigma_\mu$ where $\sigma_\mu$ now are generators of the Lie algebra
$\aso(3)\cong \asu(2)$. These solutions correspond to fuzzy
two-spheres; they form stable solutions of the IKKT matrix model deformed by
a Myers term if the coupling $\gamma$ is sufficiently
large~\cite{Azuma:2004zq}. This is consistent with the stability we
find in our model; the Hessian $\frac{\delta^2 S}{\delta A_\mu^a\,
  \delta A_\nu^b}$, $a,b=1,2,3$ is a $9\times 9$ matrix with
eigenvalues $(0,1)\, \gamma^2$ of respective multiplicities $(3,6)$. The three flat directions correspond to rotations of the fuzzy $S^2$.

\section{Interpretation as noncommutative field theories}

\subsection{General considerations}

Consider a solution to the classical equations of motion of the IKKT
matrix model corresponding to a noncommutative space. It is well-known
that the expansion of those scalar fields in this background which
acquire non-trivial values in this solution yields the action of
(supersymmetric) Yang-Mills theory on that noncommutative space. This
expansion is of the general form $X^I=x^I+Y^I$, where
$x^I$ corresponds to the classical solution and take values in a
certain Lie algebra $\frg$. The fluctuations around the noncommutative
background $Y^I$ are then taken to be valued in the tensor product of the universal enveloping algebra $\CU(\frg)$ and the gauge algebra.

It is tempting to apply the same reasoning to the quantum
Nambu-Poisson geometries arising in our model. As observed
in~\cite{DeBellis:2010pf}, however, the 3-Lie algebra structure cannot
be extended to a 3-Lie algebra structure on the whole universal enveloping
algebra. In~\cite{DeBellis:2010pf}, we concluded that the 3-Lie
algebra structure appears only at linear level. This makes a direct
expansion as above impossible. Instead, we have to choose an explicit
form of the 3-bracket on the universal enveloping algebra of the 3-Lie
algebra, which then turns into the 3-bracket at linear level.

To extend the 3-bracket of a 3-Lie algebra to its universal enveloping
algebra, one can either give up the fundamental identity or total
antisymmetry of the 3-Lie bracket beyond linear order. In the latter
case, one arrives either at the generalized 3-Lie algebras
of~\cite{Cherkis:2008qr} or the Hermitian 3-Lie algebras
of~\cite{Bagger:2008se} giving a matrix model of the ABJM theory. As
stated before, we are interested in descriptions of quantized
Nambu-Poisson manifolds as described in~\cite{DeBellis:2010pf}. For
that reason, we will choose to work with a totally antisymmetric
operator product and give up the fundamental identity.

\subsection{Structures on the universal enveloping algebra}

Consider a 3-Lie algebra $\CA$. We define its universal enveloping
algebra $\CU(\CA)$~\cite{DeBellis:2010pf} as the quotient of the tensor algebra of the
underlying vector space of $\CA$ by the two-sided ideal generated by
the relations
\begin{equation}\label{eq:factoruniversal}
 [\tau_{a_1},\tau_{a_2},\tau_{a_3}]-\sum_{i,j,k=1}^3\, \eps_{ijk}\, \tau_{a_i}\otimes \tau_{a_j}\otimes \tau_{a_k}=0~,
\end{equation}
where $\tau_a$ are generators of $\CA$. In practice, we will represent
the universal enveloping algebra in terms of finite-dimensional
matrices which will lead to the factoring by further ideals, as e.g.\
$\sum_a\, \tau_a^2=R^2$ in the case of the fuzzy three-sphere. We call the resulting algebras the factored universal enveloping algebras.

On $\CU(\CA)$, we can in principle define two distinct totally antisymmetric operator products. (Note that we have to make sure in each concrete case that this definition is really invariant on the equivalence classes defined by \eqref{eq:factoruniversal}.) The first one is defined by demanding that the bracket is totally antisymmetric and that it satisfies the generalized Leibniz rule
\begin{equation}\label{eq:3bracket1}
 [A,B,\tau_a\otimes C]=\tau_a\otimes[A,B,C]+[A,B,\tau_a]\otimes C~,
\end{equation}
where $\tau_a\in \CA$ and $A,B,C\in\CU(\CA)$. This definition ensures
that the action of the associated Lie algebra $\frg_\CA$ extends
nicely to the universal enveloping algebra $\CU(\CA)$, i.e. we have
\begin{equation}
 [\tau_a,\tau_b,\tau_{c_1}\otimes\cdots\otimes \tau_{c_l}]=
 \sum_{i=1}^l\, 
 \tau_{c_1}\otimes\cdots\otimes\tau_{c_{i-1}}\otimes
 [\tau_a,\tau_b,\tau_{c_i}] \otimes\tau_{c_{i+1}}\otimes
 \cdots\otimes\tau_{c_l} \ .
\end{equation}
As this 3-bracket is defined recursively, it is rather difficult to
handle.

The second option is the simpler definition of 
\begin{equation}
 [A_1,A_2,A_3]:=\eps_{ijk}\, A_i\otimes A_j\otimes A_k~,
\end{equation}
which evidently reduces to the 3-Lie algebra bracket for elements
$A_i$ of $\CU(\CA)$ which are also elements of $\CA$. By using this
product, we essentially ignore the associative action of the Lie
algebra $\frg_\CA$. However, we found in~\cite{DeBellis:2010pf} that
it is this operator product that is most suitable for e.g. the description of $\FR^3_\lambda$.

Consider now a solution $x^I\in\CA$ to the equations of motion \eqref{eq:eoms3M}. We take $x^I$ as a background and expand around it as
\begin{equation}\label{eq:expansion}
 X^I=x^I+Y^I~,
\end{equation}
where $Y^I$ is valued in $\CU(\CA)$. To plug this expansion into the action \eqref{eq:action3M}, we need an extension of the metric on the 3-Lie algebra to the universal enveloping algebra. In the concrete examples we will study in the following, such a metric will always appear naturally. We will now interpret the result of substituting the expansion \eqref{eq:expansion} into the action \eqref{eq:action3M} extended to the universal enveloping algebra $\CU(\CA)$ as a field theory on the noncommutative space described by $\CU(\CA)$.

\subsection[Field theory on fuzzy S**3]{Field theory on fuzzy $S^3$}

Recall that the construction of fuzzy $S^3$ makes use of a subalgebra
of endomorphisms on the Hilbert space of fuzzy $S^4$. The latter space
is obtained by embedding $S^4$ into $\CPP^3$. The algebra of
quantized functions on $\CPP^3$ is given by $N\times N$ Hermitian matrices
with $N=\frac{(3+k)!}{3!\, k!}$ and $k\in\NN$. From this construction
it is clear that there is an embedding of a reduced universal
enveloping algebra of the Clifford algebra $C\ell(\FR^5)$ in $\au(N)$,
and the same is then true for the correspondingly reduced universal
enveloping algebra of $C\ell(\FR^4)$. As discussed
in~\cite{DeBellis:2010pf}, one possible totally antisymmetric operator
product is here a totally antisymmetric matrix product combined with
an external matrix as
\begin{equation}\label{eq:3bracket}
 \begin{aligned}
 {}[A_1,A_2,A_3] := & \ [A_1,A_2,A_3,\gamma_5]\\[4pt]
 = & \ \eps_{ijk}\, (A_i\, A_j\, A_k\, \gamma_5-A_i\, A_j\, \gamma_5\, A_k+A_i\, \gamma_5\, A_j\, A_k-\gamma_5\, A_i\, A_j\, A_k)~.
 \end{aligned}
\end{equation}

Nevertheless, here we prefer to use a bracket constructed recursively as in \eqref{eq:3bracket1}. As a scalar product on $\au(N)$, we will use the standard Hilbert-Schmidt norm. The noncommutative field theory then contains the term
\begin{equation}\label{eq:S3kinetic}
 \tr\big([x^i,x^j,Y^k]\, [x^i,x^j,Y^k] \big)~.
\end{equation}
As our recursively defined 3-bracket \eqref{eq:3bracket1} allows us to
lift the action of the associated Lie algebra $\frg_{A_4}$ to the
universal enveloping algebra, this term reproduces the desired kinetic
term in our matrix model. More explicitly, the generators $x^i$ are
mapped to the generators of $C\ell(\FR^4)$ embedded into $\au(N)$ by a
homomorphism $\rho$, and this term reads
\begin{equation}
\tr\big([x^i,x^j,Y^k]\, [x^i,x^j,Y^k] \big)=\tr\Big(
\big[\rho(\gamma^{ij}),T(Y^k) \big]\, \big[\rho(\gamma^{ij}),T(Y^k)
\big] \Big)~,
\end{equation}
where $T(Y^k)$ is the polynomial $Y^k$ in $x^i$ with the replacements
$x^1\leftrightarrow x^4$ and $x^2\leftrightarrow x^3$.

Besides the kinetic term \eqref{eq:S3kinetic}, there is a mass and potential terms,
\begin{equation}
 \tr(Y^i\, Y^i)~,~~~~\eps_{ijkl}\,
 \tr\big([Y^i,Y^j,Y^k]\,Y^l\big)~,~~~~\tr\big([Y^i,Y^j,Y^k]\, [Y^i,Y^j,Y^k]
 \big)~,
\end{equation}
and the constant terms (with dimensionful prefactors)
\begin{equation}
\begin{aligned}
 \tr(x^i\, x^i)= R^2~,~~~~\eps_{ijkl}\, \tr\big([x^i,x^j,x^k]\, x^l
 \big)= 4! \, R^4~,~~~~\tr\big([x^i,x^j,x^k]\, [x^i,x^j,x^k] \big)=
 4!\, R^6~.
\end{aligned}
\end{equation}
There are also momentum-dependent terms
\begin{equation}
\eps_{ijkl}\, \tr\big([x^i,x^j,Y^k]\,
Y^l\big)~,~~~~\tr\big([x^i,x^j,Y^k]\,
[x^i,Y^j,x^k]\big)~,~~~~\tr\big([x^i,x^j,Y^k]\, [Y^i,Y^j,Y^k]\big)~.
\end{equation}
There are further terms appearing in the action, but they do not allow
for an immediate interpretation.

The momentum-dependent terms are
reminiscent of those which occur in recent proposals for
renormalizable noncommutative gauge theories, which are modifications
of the standard noncommutative Yang-Mills theory that eliminate the usual problems
associated with UV/IR mixing; see~\cite{Blaschke:2010kw} for a recent review. It would be interesting to investigate the behaviour
of our induced quantum gauge theories in more detail along these lines.
In the strong coupling limit, it is clear that the potential terms
reduce appropriately to the potential terms of the usual gauge theory
on fuzzy $S^2$. The inner derivations of $\frg_{A_4}\cong \aso(4)$ in the representation $\au(N)$, $\rho(\gamma^{ij})$, are reduced to representations of $\asu(2)$ given by $\rho(\gamma^{i4})$ with
\begin{equation}
 \big[\rho(\gamma^{i4}),\rho(\gamma^{j4}) \big]= \eps^{ijk}\,
 \rho(\gamma^{k4})\, \gamma_5~.
\end{equation}

\subsection[Field theory on R**3-lambda]{Field theory on $\FR^3_\lambda$}

The space $\FR^3_\lambda$ arising from the Nambu-Heisenberg 3-Lie
algebra $\CA_{\rm NH}$ can be considered as a discrete foliation of
$\FR^3$ by fuzzy two-spheres. Recall that the algebra of endomorphisms
of $\FR^3_\lambda$ is given by $\CE:=\bigoplus_{k\in\NN}\, \au(k)$,
where each integer $k$ corresponds to a fuzzy sphere. Thus we are
looking for a representation of the universal enveloping algebra of
the Nambu-Heisenberg 3-Lie algebra $\CA_{\rm NH}$ on $\CE$. As shown
in~\cite{DeBellis:2010pf}, we can use the totally antisymmetric matrix
product for this purpose, i.e.
\begin{align}\label{totallyantisymmetricbracket}
[A,B,C]=\tfrac{1}{6}\, \big(A\, [B,C]+B\, [C,A]+C\, [A,B] \big)~,~~~~A,B,C\in\CE~,
\end{align}
and at linear level, where the generators $\tau_i$ of $\CA_{\rm NH}$
correspond to the endomorphisms in $\CE$ describing linear coordinate
functions on all the fuzzy spheres, this product reproduces the
Nambu-Heisenberg 3-Lie algebra. It is clear that a central element of
the Lie algebra will not be central in the 3-Lie algebra. In
particular, one has
\begin{equation}
[\unit,A,B]=\alpha\, [A,B]~,~~~~\alpha \in \mathbb{C}^\times~.
\end{equation}
This issue was already discussed in~\cite{DeBellis:2010pf}.

The expansion of the action around the background solution $x^i$
satisfying the relations of the Nambu-Heisenberg 3-Lie algebra $\CA_{\rm NH}$ is given in terms of
\begin{equation}
 X^i=x^i+Y^i=\bigoplus_{k=1}^\infty\, \left(\rho_k(\sigma^i)+Y^i_k\right)~,
\end{equation}
with the generators $\sigma^i$ given in the $k$-dimensional irreducible
representation $\rho_k$ of $\asu(2)$ and $Y^i_k$ are elements of $\au(k)$ on
which these representations act. It follows, in particular, that the
action will split into a sum of separate actions on each fuzzy
sphere. While the expected kinetic terms on each sphere corresponding
to the second Casimir operator of $\asu(2)$ in the representation formed by $\au(k)$ is contained in the actions, the terms corresponding to the radial derivative (which would have to be discrete, cf.~\cite{Hammou:2001cc}) is absent. The expanded field theory will therefore not yield the expected noncommutative gauge theory.

\subsection{Field theory on more general backgrounds}

We saw above that one of the major problems in obtaining
noncommutative field theories from the 3-Lie algebra model is arriving
at the appropriate kinetic terms in the action. Another approach to
this problem would be to consider solutions with non-trivial gauge potential
$A_\mu$, from which the appropriate kinetic terms are constructed or
at least complemented. This strategy would certainly work for
$\FR^3_\lambda$. For spaces of dimension larger than three, this is
however much less clear. Moreover, this approach further deviates from
the original philosophy of constructing noncommutative gauge theories
from zero-dimensional field theories. We therefore refrain from going
into any further details.

\section[osp(1|32)-invariance]{$\aosp(1|32)$-invariance}

Some time ago, the superalgebra $\aosp(1|32)$ was suggested to be the symmetry algebra underlying M-theory. This led Smolin to study cubic matrix models with matrices taking values in $\frg\otimes \aosp(1|32)$, where $\frg$ is a gauge algebra \cite{Smolin:2000kc}. In the following, we will point out the close relationship between our 3-Lie algebra reduced model and these cubic supermatrix models.

We start from the observation that the BLG theory can be rewritten using fields taking values in the Clifford algebra of $\FR^{1,10}$ as partly done e.g. in \cite{Cherkis:2008ha}. Let $\Gamma=(\Gamma_\mu,\Gamma_I)$ denote the generators of this Clifford algebra, satisfying
\begin{equation}
 \{\Gamma_\mu,\Gamma_\nu\}=\eta_{\mu\nu}~,~~~~\{\Gamma_\mu,\Gamma_I\}=0~,~~~~\{\Gamma_I,\Gamma_J\}=\delta_{IJ}~.
\end{equation}
We combine the components of the gauge potential $A_\mu$ into the Clifford algebra valued object $A:=\Gamma^\mu\, A_\mu$ and similarly, instead of the scalar fields $X^I$, we work with $X=\Gamma_I\, X^I$. We also introduce the derivative operator $\dpar :=\Gamma^\mu\, \dpar_\mu$, such that the Dirac operator is $\slasha{\nabla}:=\Gamma^\mu\,\nabla_\mu=\dpar+A$.

The BLG Lagrangian associated to a metric 3-Lie algebra $\CA$ with associated Lie algebra $\frg_\CA$ can now be written as 
\begin{equation}\label{eq:BLGClifford}
\begin{aligned}
\CL_{\mathrm{BLG}}\ =\ &\tfrac{1}{2}\, T\big((\slasha{\nabla} X,\slasha{\nabla} X)\big)+\tfrac{\di}{2}\, (\bar{\Psi},\slasha{\nabla}\Psi) +\tfrac{\di}{4}\, \big(\bar{\Psi},[X,X,\Psi]\big)\\
&+\tfrac{1}{12}\, \tfrac{1}{6}\, T\Big(\big([X,X,X],[X,X,X]\big)\Big) -\tfrac{1}{2}\, T\Big(\Gamma_{012}\blbr A,\dpar A+\tfrac{1}{3}\, [A,A]\brbr\Big)~.
\end{aligned}
\end{equation}
Here $T(-)=\tfrac{1}{32}\, \tr_{\CC}(-)$ is the properly normalized trace over the Clifford algebra $\CC=C\ell(\FR^{1,10})$. The supersymmetry transformations are given by
\begin{equation}
  \delta_\eps X \ =\  \di\, \Gamma_I\,\left(\bar{\eps}\, \Gamma^I\Psi\right)~,~~~~
  \delta_\eps \Psi\ =\ \slasha{\nabla} X\, \eps-\tfrac{1}{6}\, [X,X,X]\, \eps~,~~~~
  \delta_\eps A\ =\ \di\,\Gamma^\mu\, \big(\bar{\eps}\,\Gamma_\mu (X\wedge\Psi)\big)~.
\end{equation}

It is evident that the same rewriting of the action works in our
dimensionally reduced model \eqref{eq:action3M}. We note that a subset
of the Clifford algebra $C\ell(\FR^{1,10})$ coincides with the bosonic
part of the superalgebra $\aosp(1|32)$, cf.\
\cite{Smolin:2000kc}. Recall that $\aosp(1|32)$
is defined in terms of $33\times33$ supermatrices
\begin{equation}
\left(\begin{array}{cc|c}
A& B & \psi\\ C& -A^\top & \phi \\ \hline \phi^\top& -\psi^\top & 0
\end{array}\right) \ ,
\end{equation}
where $A,B,C$ are $16\times16$ bosonic matrices and $\psi,\phi$ are
$16$-component Majorana spinors. In particular, when using the conventions
\begin{equation}
 \begin{aligned}
  \Gamma^0=\left(\begin{array}{cc|c} 0 & -\unit & 0 \\ \unit & 0 & 0 \\ \hline 0 & 0 & 0\end{array}\right)~,~~~~\Gamma^1=\left(\begin{array}{cc|c} 0 & \unit & 0 \\ \unit & 0 & 0 \\ \hline 0 & 0 & 0\end{array}\right)~,~~~~\Gamma^{m+1}=\left(\begin{array}{cc|c} \gamma^m & 0 & 0 \\ 0 & -\gamma^m & 0 \\ \hline 0 & 0 & 0\end{array}\right)~,
 \end{aligned}
\end{equation}
where $\gamma^m$, $m=1,\ldots,9$ are $16\times 16$ symmetric real generators of $C\ell(\FR^9)$, the matrices
\begin{equation}
 \Gamma^I~,~~~~\Gamma^\mu\, \Gamma_{012}~,~~~~\Gamma^{IJ}
\end{equation}
are elements of $\aosp(1|32)$. Moreover, because of the chirality condition $\Gamma_{012}\Psi=-\Psi$, we can write the action of the 3-Lie algebra reduced model in the form
\begin{equation}\label{eq:3LMClifford}
\begin{aligned}
 S_{\mathrm{3LAM}}\ =\ &\tfrac{1}{2}\, T\big((\tilde{A}\, X,\tilde{A}\, X)\big)-\tfrac{\di}{2}\,(\bar{\Psi},\tilde{A}\, \Psi) +\tfrac{\di}{4}\, \big(\bar{\Psi},[X,X,\Psi] \big)\\
&+\tfrac{1}{12}\, \tfrac{1}{6}\, T\Big(\big([X,X,X],[X,X,X]\big)\Big) +\tfrac{1}{6}\, T\Big(\blbr \tilde{A},[\tilde{A},\tilde{A}]\brbr\Big)~,
\end{aligned}
\end{equation}
where $\tilde{A}=A\, \Gamma_{012}$. The action \eqref{eq:3LMClifford} can thus be interpreted in terms products of supermatrices in $\aosp(1|32)$ and the trace operation $T$ can be incorporated as a supertrace. The reduced model has therefore manifest $\aosp(1|32)$-invariance. Instead of developing this symmetry in more detail, let us examine how closely related the action $S_{\mathrm{3LAM}}$ is to the cubic matrix model of \cite{Smolin:2000kc}.

The dichotomy of the matter and gauge fields in our 3-Lie algebra model forces us to introduce a more complicated gauge structure than usual. We will work with a matrix $M$ taking values in $\aosp(1|32)\otimes \CV$, where $\CV:=\CA\oplus\frg_\CA$. Moreover, we define a triple bracket $\CT_3:\CV\otimes\CV\otimes\CV\rightarrow \FC$ as the cyclic product
\begin{equation}
 \CT_3(A,B,C):=\left\{\begin{array}{ll}
                     \tfrac{1}{6}\, \blbr A,[B,C] \brbr & \mbox{for}~A,B,C\in\frg_\CA \\
		     \tfrac{\di}{6}\, (A,B\, C) & \mbox{for}~A,C\in\CA
                     \ ,~B\in\frg_\CA \\
		     \tfrac{\di}{6}\, (C,A\, B) & \mbox{for}~B,C\in\CA
                     \ ,~A\in\frg_\CA \\
		     \tfrac{\di}{6}\, (B,C\, A) & \mbox{for}~A,B\in\CA
                     \ ,~C\in\frg_\CA\\
		     0 & \mbox{otherwise}
                    \end{array}\right.~~.
\end{equation}
With this definition, the cubic supermatrix model
\begin{equation}
 S_{\rm CSM}=\str\big(\CT_3(M,M,M)\big)~,
\end{equation}
with $\str(-)$ the supertrace and the matrix
\begin{equation}
 M=\left(\begin{array}{c|c}
\tilde{A}-\tfrac{1}{2}\, D(X,X)& \Psi\\ \hline \bar{\Psi} & 0
\end{array}\right) \ ,
\end{equation}
reproduces the Chern-Simons part and the fermionic part of \eqref{eq:3LMClifford}. Similarly to \cite{Azuma:2001re}, one could therefore argue that after introducing a mass-like term, nonperturbative effects would induce the bosonic part, thus completing the action. 

Here we will follow a different approach. We introduce the cyclic quadruple bracket $T_4:\CV\otimes\CV\otimes\CV\otimes\CV \rightarrow \FC$ with
\begin{equation}
 \CT_4(A,B,C,D):=\left\{\begin{array}{ll}
                     \tfrac{1}{4}\, (A\, B,C\, D) &
                     \mbox{for}~A,C\in\frg_\CA \ ,~B,D\in\CA \\
		     \tfrac{1}{4}\, (B\, C,D\, A) &
                     \mbox{for}~B,D\in\frg_\CA \ ,~A,C\in\CA \\
		     0 & \mbox{otherwise}
                    \end{array}\right.
\end{equation}
and consider the action
\begin{equation}
 S_{\rm CSM}=\str\big(\CT_3(M,M,M)\big)+\str\big(\CT_4(M,M,M,M)\big)
\end{equation}
with
\begin{equation}
 M=\left(\begin{array}{c|c}
\tilde{A}-\tfrac{1}{2}\, D(X,X)+X & \Psi\\ \hline \bar{\Psi} & 0
\end{array}\right)~.
\end{equation}
The additional term reproduces the kinetic terms for the bosonic matter fields as well as the bosonic matter potential. The potential appears with the right sign but with a different prefactor from the one appearing in \eqref{eq:3LMClifford}.
It would be interesting to study the relation between our 3-Lie algebra reduced models and the cubic matrix models of Smolin in more detail. This, however, is beyond the scope of the present paper.

\acknowledgements

We are grateful to C.-S.~Chu, J.-H.~Park and H.~Steinacker for enlightening discussions. This work was supported by grant ST/G000514/1 ``String Theory
Scotland'' from the UK Science and Technology Facilities Council. The
work of CS was supported by a Career Acceleration Fellowship from the
UK Engineering and Physical Sciences Research Council.

\appendices

\subsection{Generalized 3-Lie algebras}

Recall that a {\em 3-Lie algebra}~\cite{Filippov:1985aa} is a vector
space $\CA$ endowed with a totally antisymmetric, trilinear map
$[-,-,-]:\CA\wedge\CA \wedge\CA \rightarrow \CA$ which satisfies the {\em fundamental identity}
\begin{equation}
\big[a,b,[x,y,z]\big]=\big[[a,b,x],y,z\big]+\big[x,[a,b,y],z\big]+\big[x,y,[a,b,z]\big]
\end{equation}
for all $a,b,x,y,z\in\CA$. We can endow $\CA$ with an invariant symmetric bilinear form, i.e. a map $(-,-):\CA\odot\CA\rightarrow \FC$ satisfying the {\em compatibility condition}
\begin{equation}
 \big([a,b,x],y\big)+\big(x,[a,b,y]\big)=0
\end{equation}
for all $a,b,x,y\in\CA$. If this form is non-degenerate, then it defines a {\em metric 3-Lie algebra}.

A 3-Lie algebra $\CA$ comes with an {\em associated Lie algebra}
$\frg_\CA$ of inner derivations, which consists of the span of the linear maps $D(a,b):\CA\rightarrow \CA$, $a,b\in\CA$, defined via 
\begin{equation}
 D(a,b)x=[a,b,x]~,~~~~x\in \CA~.
\end{equation}
The invariant symmetric bilinear form on $\CA$ induces an invariant
symmetric bilinear form on $\frg_\CA$ defined through
\begin{equation}
 \blbr D(a,b),D(c,d)\brbr=\big([a,b,c],d\big)~,~~~~a,b,c,d\in\CA~.
\end{equation}
This inner product is different from the Cartan-Killing form. 

The most important example of a metric 3-Lie algebra is $\CA=A_4$. It is given by the vector space $\FR^4$ with standard basis $(e_1,e_2,e_3,e_4)$, endowed with the 3-bracket and invariant form
\begin{equation}
 [e_i,e_j,e_k]=\eps_{ijkl}\, e_l \eand (e_i,e_j)=\delta_{ij}~.
\end{equation}
Its associated Lie algebra is $\frg_{A_4}=\aso(4)\cong\asu(2)\oplus \asu(2)$, and the alternative invariant form $\lbr-,-\rbr$ on $\frg_{A_4}$ is of split signature.

A {\em generalized 3-Lie algebra}~\cite{Cherkis:2008qr} is a vector space $\CA$ endowed with a trilinear map which is antisymmetric only in its first two slots but still satisfies the fundamental identity. Contrary to~\cite{Cherkis:2008qr}, we allow for invariant bilinear forms on $\CA$ which are not positive definite. Just as in the case of 3-Lie algebras, these generalized 3-Lie algebras come with an associated Lie algebra $\frg_\CA$ possessing an alternative invariant symmetric bilinear form.
Generalized 3-Lie algebras contain 3-Lie algebras as special cases and
provide a natural extension of them~\cite{deMedeiros:2008zh}. They
also contain families which can be parameterized by an integer
$N$~\cite{Cherkis:2008ha}, similarly e.g. to the families
$\mathfrak{o}(N)$ and $\fru(N)$ of Lie algebras.

\pdfbookmark[1]{References}{refs}\label{refs}

\bibliography{bigone}
\bibliographystyle{latexeu}
\end{document}